\documentclass[twocolumn,showpacs,pre,superscriptaddress]{revtex4}
\usepackage{graphicx}
\usepackage{color}
\usepackage{amssymb,amsfonts,amsmath}
\usepackage{natbib}

\begin{document}

\title{Drift-induced Benjamin-Feir instabilities}
\author{Francesca Di Patti} \affiliation{Universit\`{a} degli Studi di Firenze, Dipartimento di Fisica e Astronomia, CSDC and INFN Sezione di Firenze, via G. Sansone 1, 50019 Sesto Fiorentino, Italia}
\author{Duccio Fanelli} \affiliation{Universit\`{a} degli Studi di Firenze, Dipartimento di Fisica e Astronomia, CSDC and INFN Sezione di Firenze, via G. Sansone 1, 50019 Sesto Fiorentino, Italia}
\author{Timoteo Carletti}\affiliation{Department of Mathematics and Namur Center for Complex Systems - naXys, University of Namur, rempart de la Vierge 8, B 5000 Namur, Belgium}

\date{\today} 

\begin{abstract}
A modified version of the Ginzburg-Landau equation is introduced which accounts for  asymmetric couplings between neighbors sites on a one-dimensional lattice, with periodic boundary conditions. The drift term which reflects the imposed microscopic asymmetry seeds a generalized class of instabilities, reminiscent of the Benjamin-Feir type. The uniformly synchronized solution is spontaneously destabilized outside the region of parameters classically associated to the Benjamin-Feir instability,  upon  injection of a non homogeneous perturbation. The ensuing patterns can be of the traveling wave type or display a patchy, colorful mosaic for the modulus of the complex oscillators amplitude. 
\end{abstract}

\pacs{05.45.Xt,  82.40.Ck, 89.75.Kd}

\maketitle
The spontaneous ability of spatially extended systems to self-organize in space and time is proverbial and has been raised to paradigm in modern science \cite{pikovsky,turing}. Collective behaviors are widespread in nature and mirror, at the macroscopic level, the microscopic interactions at play among elementary constituents.  Convection instabilities in fluid dynamics, weak turbulences and defects  are representative examples that emblematize the remarkable capacity of assorted physical systems to yield coherent dynamics \cite{drazin}. Rhythms production and the brain functions are prototypical illustrations drawn from biology \cite{goldbeter,wyller07}, insect swarms and fish schools refer instead to ecological applications \cite{murray2}. The degree of instinctive and unsupervised coordination which instigates the bottom-up cascade towards self-regulated patterns is however universal and, as such, has been invoked in many other fields \cite{nakao2010,nakao2014,asllani2013,dipatti}, ranging from chemistry \cite{belousov,zhab} to economy, via technology \cite{strogatzBook}. Instabilities triggered by random fluctuations are often patterns precursors. The imposed perturbation shakes e.g. an homogeneous equilibrium, seeding a resonant amplification mechanism that eventually materializes in magnificent patchy motifs, characterized by a vast gallery of shapes and geometries. Exploring  possible routes to pattern formation, and unraveling novel avenues to symmetry breaking instability, is hence a challenge of both fundamental and applied importance.
 
In the so-called modulational instability deviations from a periodic waveform are reinforced by nonlinearity, leading to  spectral-sidebands and the breakup of the waveform into a train of pulses \cite{benjHasselmann,agrawal}. The phenomenon was first conceptualized for periodic surface gravity waves (Stokes waves) on deep water by Benjamin and Feir, in 1967 \cite{benjamin}, and for this reason is customarily referred to as the Benjamin-Feir (BF) instability.  The BF instability has been later on discussed \cite{diprima, kuramotoBook} in the context of the Complex Ginzburg Landau equation (CGLE), a quintessential model for non linear physics, whose applications range from superconductivity, superfluidity and Bose-Einstein condensation to liquid crystals and strings in field theory \cite{reviewGLE}.  In this Letter, we will revisit the BF instability in the framework of the CGLE, modified with the inclusion of a drift term. This latter is rigorously derived from a stochastic description of the microscopic coupling between adjacent oscillators.  As we shall prove in the following, generalized BF instabilities occur, stimulated by the drift, outside the region of parameters for which the classical BF instability is manifested. This observation,  grounded on a detailed mathematical theory, contributes to considerably enrich the landscape of known instabilities, along a direction of investigation that can be experimentally substantiated.

Let us start by considering an ensemble made of $N$ non linear oscillators and label with $W_i$ their associated complex amplitude, where $i=1,...,N$.  We shall hereafter assume that each individual oscillator obeys to a CGLE. Moreover, the oscillators are mutually coupled via a diffusive-like interaction that is mathematically epitomized in terms of a discrete Laplacian operator.  Concretely,  imagine that oscillators are coupled to nearest neighbors only,  on a one dimensional lattice with periodic boundaries. The connection can be made directed by assigning different probabilities of pair interactions between adjacent sites: $a$ refers to the probability that links $i$ to $i+1$, while $b$ stands for the probability of nodes $i$ and $i-1$  to communicate.  The obvious constraint $a+b=1$ holds true. Under these premises, the discrete Laplacian operator $\mathbf{\Delta}$  results in a circulant matrix with three non trivial entries per row, namely $\Delta_{ii}=-(a+b)=-1$, $\Delta_{i,i+1}=b$ and $\Delta_{i,i-1}=a$. The action of the operator $\mathbf{\Delta}$ on the complex amplitude value $W_i$ is explicitly given by:
\begin{multline*}
\sum_{k=1}^N \Delta_{ij}W_j = \frac{(a+b)\left (\delta x \right )^2}{2}  \frac{W_{j-1} -2 W_j +W_{j+1} }{\left (\delta x \right )^2} \\
+ \frac{(b-a)\delta x }{2} \frac{W_{j+1} -W_{j-1} }{\delta x}
\end{multline*} 
where an artificial rescaling with $\delta x$, the lattice spacing, has been introduced. As expected, two contributions can be highlighted: a symmetric diffusion part and a drift term. This latter vanishes when $a=b=1/2$, i.e. when the probability of interactions of $i$ with, respectively, $i+1 $ and $i-1$ is assumed identical.   Performing the thermodynamic limit $N \rightarrow \infty$ is equivalent to operating in the continuum limit  $\delta x \rightarrow 0$. The discrete variable $W_i (t)$ is hence mapped onto $W(x,t)$, a time dependent complex function defined on a continuum, one dimensional, spatial support  $x$. The above reasoning yields the following complex  Ginzburg Landau equation with drift (CGLED) for the evolution of the complex amplitude 
$W(x,t)$:
\begin{equation}\label{eq:GLdrift}
\partial_t W =  W - (1+i c_2) \vert W \vert ^2 W +  K (1+i c_1) ( D \partial_x^2 W +v \partial_x W)
\end{equation}
where $D=\lim_{\delta x \rightarrow 0 }(a+b)/2 \left (\delta x \right )^2$ and $v=\lim_{\delta x \rightarrow 0 }(b-a) \delta x$ are the diffusion constant and the velocity, as usually defined.  Here, $K$ is a constant parameter which modulates the strength of the coupling.  In the following, and without losing generality, $K$ is set equal one. A comment is mandatory at this point. Moving from a self-consistent microscopic description of the inspected processes, one ends up with a standard CGLE modified by the inclusion of an additional drift operator, multiplied by the {\it effective} (complex) velocity $v' =(1+i c_1) v$. This is at variance with other models investigated in the literature \cite{cross}, which assumed dealing with an isolated heuristic term $v \partial_x W$, on the right hand side of the CGLE. This latter modification proves unessential, as concerns the onset of the instability. In fact, the term $v \partial_x W$ can be formally removed by performing a change of variable to the comoving reference frame. At odds with this conclusion, the complex coefficient $(1+i c_1)$,  multiplying the scalar velocity $v$ in Eq. (\ref{eq:GLdrift}), mixes the real and imaginary components of $W$, an apparently innocent step which however significantly alters the emerging dynamics for the modified CGLE. As a matter of fact, the inherited degree of complexity is ultimately responsible for the generalized instability that we shall henceforth report.   

The CGLED \eqref{eq:GLdrift} admits a family of traveling wave (TW) solutions of the type :
\begin{equation}\label{eq:TW}
W_{TW}= R e^{i \omega t+i k x}
\end{equation}
where $R^2=1-k^2D-k v c_1$ and $\omega=-R^2 c_2+k v-k^2 D c_1$. Particularly interesting is the choice $k=0$, which returns $R=1$ and $\omega=-c_2$. Hence, $e^{-i c_2 t}$ is an homogenous solution of the CGLED. This latter is here termed the limit cycle (LC) solution, as it 
results from a uniform, fully synchronized, replica of the periodic orbit displayed by the system in its a-spatial ($K=0$) version. 

To shed light onto the dynamics of the Ginzburg-Landau equation, modified with the inclusion of a complex drift term, we begin by determining the stability of the TW solutions (\ref{eq:TW}). To this end we set:
\begin{equation}\label{eq:perturbTW}
W=W_{TW} \left ( 1+ a_+ (t) e^{i Q x} + a_-(t) e^{-i Q x} \right )
\end{equation}
and substitute Eq. \eqref{eq:TW} and Eq. \eqref{eq:perturbTW} into Eq. \eqref{eq:GLdrift}. At the linear order of approximation in $a_+$ and $a_-$ one obtains the following system of differential equations
\begin{equation*}
\frac{d}{dt}
\left (
\begin{matrix}
a_+\\
\bar{a}_-
\end{matrix}
\right)
=
{\bf J}
\left (
\begin{matrix}
a_+\\
\bar{a}_-
\end{matrix}
\right)
\end{equation*}
where the bar stands for the complex conjugate and the entries of matrix ${\bf J}$ read 
\begin{equation}\label{eq:matrixJ}
\begin{array}{l}
J_{11} = - Q D (Q+2 k )(1 + i c_1) -(1+i c_2) R^2 \\
\phantom{fffff}+ i Q v (1 + i c_1) \\
J_{12} = \bar{J}_{21} = -(1+i c_2) R^2 \\
J_{22} = + Q D (2 k-Q )(1 - i c_1) -(1-i c_2) R^2 \\
\phantom{fffff}+ i Q v (1 - i c_1)  \quad .
\end{array}
\end{equation}
The eigenvalues $\lambda$ of matrix ${\bf J}$ determine the asymptotic fate of the perturbation: if the real part of  $\lambda$ is positive, the perturbation grows exponentially, otherwise it fades away. To proceed in the analysis 
we selected the eigenvalue with largest real part,  denoted $\lambda^+$, and Taylor expand it for small 
$Q$ to eventually get:
\begin{equation*}
\lambda^+ \simeq q_1 Q +q_2 Q^2 + \mathcal{O}(Q^3)
\end{equation*}
where 
\begin{multline*}
q_1=\left [-2 c_1 D k+  v -(-2 c_2 D k + 2 c_2 D^2 k^3 - c_1 c_2 v \right . \\
\left . + 3 c_1 c_2 D k^2 v + c_1^2 c_2 k v^2)/R^2\right] i 
\end{multline*}
is imaginary, while $q_2$ is real and reads
\begin{equation}\label{eq:q2}
q_2=a_2 k^2 +a_1 k +a_0
\end{equation}
with 
\begin{eqnarray*}
a_2 &=& D^2(3+c_1 c_2 + 2 c_2^2)/R^2\\
a_1 &=& D v ( 3 c_1+ c_1^2 c_2 + 2 c_1 c_2^2)/R^2\\
a_0  &=& \left [ c_1^2(1+c_2^2) v^2 -2 D (1 +c_1 c_2) \right]/(2 R^2) \quad .
\end{eqnarray*}

As already mentioned, the imposed perturbation gets magnified when the real part of $\lambda^+$ is  positive or, stated differently, when $q_2>0$.  Consider first the synchronized LC solution, which corresponds to setting $k=0$. If the drift is silenced, i.e. $v=0$, requiring $q_2(k=0)>0$ yields $(1+c_1 c_2)<0$: this is the classical condition for the BF instability to hold true.  The synchronous LC homogeneous configuration gets thus spoiled by any externally enforced perturbation, provided $c_1$ and $c_2$  match the prescribed condition. The non linear evolution of the perturbation materializes in beautiful and uneven patterns for the modulus of  the complex amplitude $W$. If $(1+c_1 c_2)>0$ and $v=0$, no patterns can develop from a local perturbation of the homogeneous LC solution. As we shall prove in the following, this simplified scenario should be drastically revised when $v \ne 0$. Notice that  $a_1=0$ and $a_2>0$, for $v=0$ and $(1+c_1 c_2)>0$. Hence,  modes with $k>k_c=\sqrt{|a_0|/a_2}$ are unstable to externally applied perturbations.

\begin{figure}[tb]
\begin{center}
\includegraphics[scale=0.32]{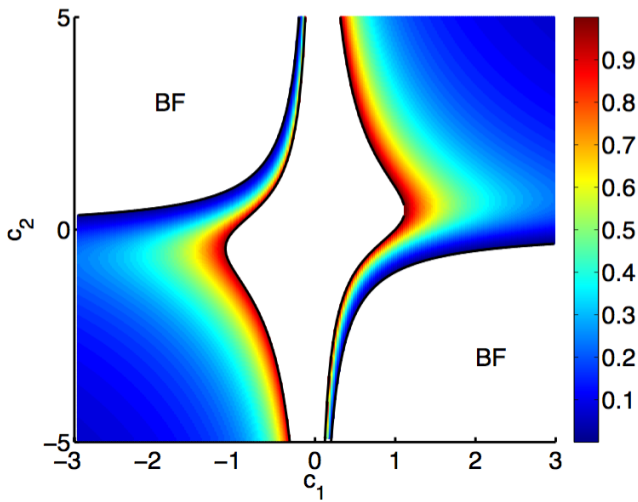}
\end{center}
\caption{The solid lines delimit the portion of the reference plane $(c_1,c_2)$ where  $0<\gamma_c < 1$. The color identifies the minimal value of $\gamma$, namely $\gamma_c$, which needs to be accommodated for to make the homogeneous LC solution unstable to externally imposed, non homogeneous, perturbation. The solid lines refer to boundaries of the standard BF instability ($1+c_1 c_2=0$) and to the edge of the drift driven instability as determined by Eq.  \eqref{edge}.
\label{fig:instabilityRegion}}
\end{figure}

Let us now turn to examining the setting with $v \ne 0$.  The homogeneous solution ($k=0$) can be made unstable under this generalized scenario, also in the region of the parameters $(1+c_1 c_2)>0$, for which the classical Benjamin-Feir instability is prevented to occur. To clarify this point, introduce $\gamma = v^2/(2 D) = (b-a)^2$. Then $q_2(k=0)=a_0>0$ provided 
\begin{equation*}
\gamma>\gamma_c = \dfrac{1+c_1 c_2}{c_1^2 \left ( 1+c_2^2 \right )}
\end{equation*}
The sought instability can hence realize provided a sufficient degree of coupling asymmetry is enforced into the model. Since $a,b \in [0, 1]$, then $\gamma_c \in [0,1]$, a constraint that limits the portion of the parameter plane $(c_1,c_2)$ where the drift-induced Benjamin-Feir instability can possibly materialize. The condition $\gamma_c>0$ is automatically satisfied since we are focusing on the restricted region of interest  $1+c_1c_2 >0$. The upper bound $\gamma_c = 1$ is reached when:  
\begin{equation}
\label{edge}
c_2=\frac{1}{2 c_1} \left [ 1 \pm \sqrt{ 5 -4 c_1^2} \right ]
\end{equation}
In Figure \ref{fig:instabilityRegion}, the critical value of  $\gamma$ is reported, with an apt color code, when scanning the plane $(c_1,c_2)$. The solid lines mark the region of interest, as outlined above.  As expected, $\gamma_c$ tends to zero when one approaches the boundary for the outbreak of the standard BF instability. 

\begin{figure}[tb]
\begin{center}
\includegraphics[scale=0.45]{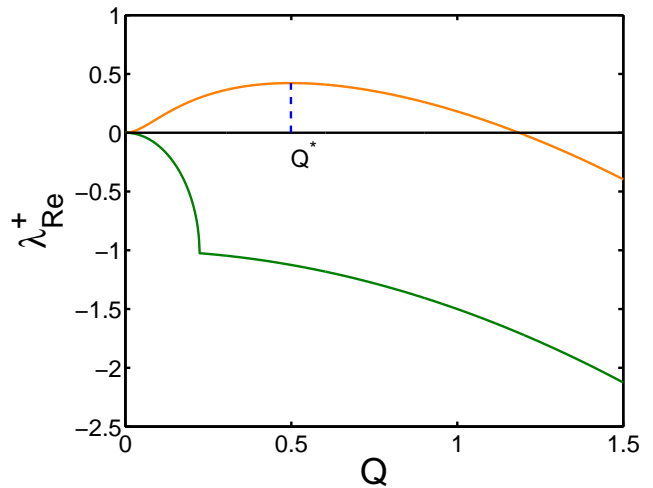} 
\end{center}
\caption{Real part of the dispersion relation (\ref{disprelLC}), $\lambda^{+}_{Re}$, plotted as function of $Q$. The curve above zero (green online) refers to $v=0$: the uniform LC solution is stable and non BF instability can occur. The dispersion relation that extends above zero (orange online) is obtained by setting $a \ne b$ so as to yield  $\gamma=\gamma_c+\Delta \gamma$, with $\Delta \gamma=0.1$. The uniform LC solution can be now destabilized. The vertical dashed blue line identifies the value $Q^*$ that yields the maximum in  $\lambda^+_{Re}$.  Here, $c_1=-2$ and  $c_2=-20$. \label{fig:relDispCL}}
\end{figure}

Summing up the system can turn unstable outside the region classically deputed to the BF instability, a phenomenon instigated by drift, the  macroscopic imprint of the endowed spatial asymmetry. To make this point transparent we go back to considering the exact expression for $\lambda^+$, as it emanates from matrix (\ref{eq:matrixJ}). For $k=0$ (and recalling that, consequently, $\omega =-c_2$ and $R=1$) one 
obtains:
\begin{multline}
\label{disprelLC}
\lambda^{+} = -1 - Q ^2 D+ i  Q v \\
+ \sqrt{1+2c_1 c_2 \left ( - Q ^2 D+ i Q  v  \right) -c_1^2 \left ( -Q ^2 D+ i  Q  v  \right) ^2 }
\end{multline}
In Fig. \ref{fig:relDispCL} the real part of the above dispersion relation ($\lambda^{+}_{Re}$) is plotted versus $Q$, the wavenumber which characterizes the perturbation, acting on the uniform LC solution ($k=0$). The values of $c_1$ and $c_2$ are set so that 
$1+c_1c_2>0$. When $v=0$  (green solid line) $\lambda^{+}_{Re}$ is always smaller (or equal for $Q=0$) than zero: no modes can be excited, and the LC keeps thus stable. The orange curve refers instead to the situation where $a \ne b$. The parameters $a$ and $b$ have been chosen so to have  $\gamma > \gamma_c (c_1,c_2)$: as predicted by the analysis carried out above, the dispersion relation lifts above zero, signaling an instability over a finite windows of $Q$. The maximum of $\lambda^{+}_{Re}$ vs. $Q$ identifies the most unstable mode $Q^*$, as illustrated in Fig. \ref{fig:relDispCL}.

A question then arises concerning the final patterns attained by the CGLED, as follows the generalized instability illustrated above and beyond the linear regime of evolution. To formulate an answer we take advantage from the observation that the dispersion relation $\lambda^+_{Re}$ displays an isolated  maximum at $Q=Q^*$. The perturbation at $Q^*$ will stand comparison with other and emerge from the sea of, self-consistently activated, modes. Imagine that solution (\ref{eq:TW}) with $k=Q^*$ proves stable, as classified by the sign of $q_2(k=Q^*)$ in Eq. \eqref{eq:q2}. Then,  one can plausibly expect that the system will asymptotically organize in a steady traveling wave, in the late non linear regime of the evolution. Contrariwise,  the mutually exclusive competition between linearly unstable modes, on the one side, and the stability conditions of traveling wave solutions, on the other, prompts more intricate, spatially inhomogeneous equilibria, for the complex amplitude $W$,  when $q_2(k=Q^*)>0$. 

\begin{figure*}[tb]
\begin{center}
\includegraphics[scale=0.55]{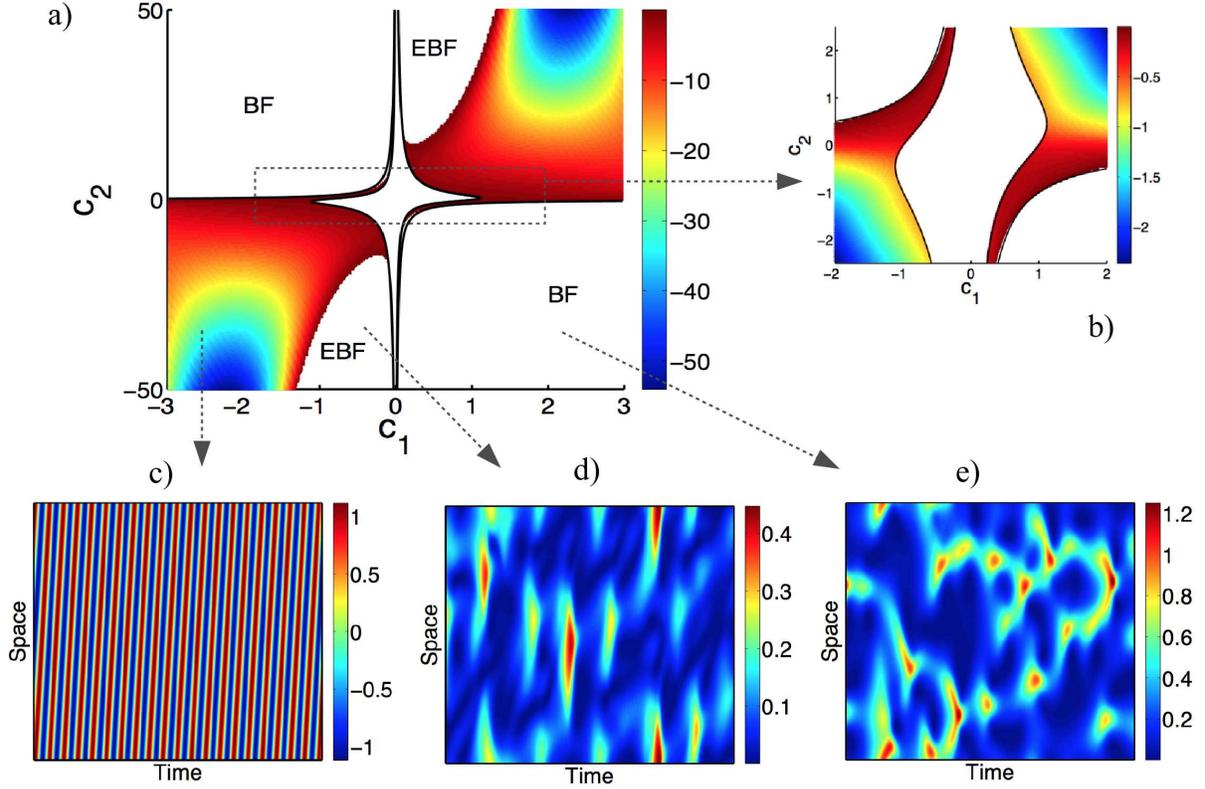} 
\end{center}
\caption{The colored regions of panel (a) mark the portions of the plane $(c_1,c_2)$ where the traveling wave $W_{TW}= R e^{i \omega t+i Q^* x}$ is stable. The color follows $q_2(Q^*)$, which, in this region, is bound to negatives values. The two regions labeled EBF are those where $q_2(Q^*)$ is found to be positive. Panel (b) is a zoom of panel (a), centered at the origin of the axes.  Panel (c) displays the real part of $W$, obtained from a  numerical solution of Eq. \eqref{eq:GLdrift}, in the plane $(t,x)$, for  $c_1=-2$, $c_2=-20$. Panel (d) shows the evolution of the modulus of the complex  amplitude $W$, as determined from a numerical integration of Eq. \eqref{eq:GLdrift} in the EBF region. The parameter are $c_1=-0.5$ and $c_2=-40$. In panel (e) the evolution of the modulus of the complex  amplitude $W$ is depicted, for a choice of the parameters which falls inside the standard region of BF instability ($c_1=1$, $c_2=-7$). In panels a), b), c) and d) $\gamma=\gamma_c+\Delta \gamma$, with $\Delta \gamma=0.1$. In panel e) $v=0$ which implies $\gamma=0$. 
\label{fig:stabilityGeneralized}}
\end{figure*}

To further develop this argument we swept through the plane $(c_1,c_2)$, by setting $\gamma=\gamma_c+\Delta \gamma$ (being $\Delta \gamma$ constant). For each pair $(c_1,c_2)$, identified the corresponding $Q^*$ and measured the associated $q_2(k=Q^*)$. The colored region in Fig.  \ref{fig:stabilityGeneralized}, panels a) and b), refers to $q_2(k=Q^*)<0$: colors reflect the measured value of $q_2$ and assume an appropriate code, as indicated in the annexed colorbar. The white regions that fall outside the  domain of classical BF instability are characterized by   $q_2(k=Q^*)>0$.  Patchy inhomogeneous patterns are expected when the system is initialized in such regions, based on the reasoning developed above. In light of the similarities that the generated patterns bear with the original BF motifs (as demonstrated below  in Fig. \ref{fig:stabilityGeneralized}, panels d) and e)), we term this regions with the acronym EBF, from Extended Benjamin-Feir instability.

In Fig. \ref{fig:stabilityGeneralized} we report the results of direct integration of the CGLED, for different choice of the parameters and starting with a random (uniformly distributed) perturbation of the homogeneous LC solution. When  $c_1$ and $c_2$ are selected from the colored region of panel a) of Fig. \ref{fig:stabilityGeneralized}, the system stabilizes on a traveling wave (see Fig. \ref{fig:stabilityGeneralized} c) ). When the parameters are  instead chosen  from the EBF regions, the amplitude of the complex amplitude W displays a complicated mosaic of beautiful patterns in the plane $(x,t)$ (see Fig. \ref{fig:stabilityGeneralized} d) ).

In conclusion, we have here introduced a modified version of the celebrated Ginzburg-Landau equation which accounts for  asymmetric couplings between neighbors sites. The drift term that descends from the imposed degree of unevenness, seeds the emergence of a generalized class of instabilities reminiscent of the Benjamin-Feir type. The uniformly synchronized solution is spontaneously destabilized, upon  injection of a non homogeneous perturbation. The ensuing patterns can be of the traveling wave type or display a colorful textures in the space and time. The conditions that discriminate between those alternatives are worked out in a rigorous mathematical framework. As a final remark, we emphasize that the theory here developed can be extended to describe the dynamics of a ensemble made of oscillators,  coupled via a direct and heterogeneous network \citep{nakao2014, asllani14, contemori}, as  we shall report elsewhere. 
\begin{acknowledgments}
This work has been supported by the program PRIN 2012 founded by the Italian Ministero dell'Istruzione, dell'Universit\`{a} e della Ricerca (MIUR). D.F. acknowledges support from the COSMOS ITN EU project. The work of T.C. presents research results of the Belgian Network DYSCO (Dynamical Systems, Control, and Optimization), funded by the Interuniversity Attraction Poles Programme, initiated by the Belgian State, Science Policy Office.
\end{acknowledgments}
\bibliography{bibliography}{}
\end{document}